\PassOptionsToPackage{unicode}{hyperref}
\PassOptionsToPackage{hyphens}{url}
\PassOptionsToPackage{dvipsnames,svgnames,x11names}{xcolor}
\documentclass[
  12pt,
  letterpaper,
]{article}

\usepackage{amsmath,amssymb}
\usepackage{setspace}
\usepackage{iftex}
\ifPDFTeX
  \usepackage[T1]{fontenc}
  \usepackage[utf8]{inputenc}
  \usepackage{textcomp} 
\else 
  \usepackage{unicode-math}
  \defaultfontfeatures{Scale=MatchLowercase}
  \defaultfontfeatures[\rmfamily]{Ligatures=TeX,Scale=1}
\fi
\usepackage{lmodern}
\ifPDFTeX\else  
\fi
\IfFileExists{upquote.sty}{\usepackage{upquote}}{}
\IfFileExists{microtype.sty}{
  \usepackage[]{microtype}
  \UseMicrotypeSet[protrusion]{basicmath} 
}{}
\makeatletter
\@ifundefined{KOMAClassName}{
  \IfFileExists{parskip.sty}{%
    \usepackage{parskip}
  }{
    \setlength{\parindent}{0pt}
    \setlength{\parskip}{6pt plus 2pt minus 1pt}}
}{
  \KOMAoptions{parskip=half}}
\makeatother
\usepackage{xcolor}
\usepackage[top=0.75in,bottom=0.75in,left=1in,right=1in]{geometry}
\makeatletter
\ifx\paragraph\undefined\else
  \let\oldparagraph\paragraph
  \renewcommand{\paragraph}{
    \@ifstar
      \xxxParagraphStar
      \xxxParagraphNoStar
  }
  \newcommand{\xxxParagraphStar}[1]{\oldparagraph*{#1}\mbox{}}
  \newcommand{\xxxParagraphNoStar}[1]{\oldparagraph{#1}\mbox{}}
\fi
\ifx\subparagraph\undefined\else
  \let\oldsubparagraph\subparagraph
  \renewcommand{\subparagraph}{
    \@ifstar
      \xxxSubParagraphStar
      \xxxSubParagraphNoStar
  }
  \newcommand{\xxxSubParagraphStar}[1]{\oldsubparagraph*{#1}\mbox{}}
  \newcommand{\xxxSubParagraphNoStar}[1]{\oldsubparagraph{#1}\mbox{}}
\fi
\makeatother

\providecommand{\tightlist}{%
  \setlength{\itemsep}{0pt}\setlength{\parskip}{0pt}}\usepackage{longtable,booktabs,array}
\usepackage{calc} 
\usepackage{etoolbox}
\makeatletter
\patchcmd\longtable{\par}{\if@noskipsec\mbox{}\fi\par}{}{}
\makeatother
\IfFileExists{footnotehyper.sty}{\usepackage{footnotehyper}}{\usepackage{footnote}}
\makesavenoteenv{longtable}
\usepackage{graphicx}
\makeatletter
\newsavebox\pandoc@box
\newcommand*\pandocbounded[1]{
  \sbox\pandoc@box{#1}%
  \Gscale@div\@tempa{\textheight}{\dimexpr\ht\pandoc@box+\dp\pandoc@box\relax}%
  \Gscale@div\@tempb{\linewidth}{\wd\pandoc@box}%
  \ifdim\@tempb\p@<\@tempa\p@\let\@tempa\@tempb\fi
  \ifdim\@tempa\p@<\p@\scalebox{\@tempa}{\usebox\pandoc@box}%
  \else\usebox{\pandoc@box}%
  \fi%
}
\def\fps@figure{htbp}
\makeatother

\usepackage{algorithm}
\usepackage{algpseudocode}
\usepackage{graphicx}
\usepackage{scrlayer-scrpage}
\usepackage{etoolbox}
\usepackage{setspace}
\clearpairofpagestyles
\ofoot*{\pagemark}
\makeatletter
\@ifpackageloaded{caption}{}{\usepackage{caption}}
\AtBeginDocument{%
\ifdefined\contentsname
  \renewcommand*\contentsname{Table of contents}
\else
  \newcommand\contentsname{Table of contents}
\fi
\ifdefined\listfigurename
  \renewcommand*\listfigurename{List of Figures}
\else
  \newcommand\listfigurename{List of Figures}
\fi
\ifdefined\listtablename
  \renewcommand*\listtablename{List of Tables}
\else
  \newcommand\listtablename{List of Tables}
\fi
\ifdefined\figurename
  \renewcommand*\figurename{Figure}
\else
  \newcommand\figurename{Figure}
\fi
\ifdefined\tablename
  \renewcommand*\tablename{Table}
\else
  \newcommand\tablename{Table}
\fi
}
\@ifpackageloaded{float}{}{\usepackage{float}}
\floatstyle{ruled}
\@ifundefined{c@chapter}{\newfloat{codelisting}{h}{lop}}{\newfloat{codelisting}{h}{lop}[chapter]}
\floatname{codelisting}{Listing}

\makeatother
\makeatletter
\@ifpackageloaded{caption}{}{\usepackage{caption}}
\@ifpackageloaded{subcaption}{}{\usepackage{subcaption}}
\makeatother

\usepackage[]{natbib}
\usepackage{bookmark}

\IfFileExists{xurl.sty}{\usepackage{xurl}}{} 
\hypersetup{
  colorlinks=true,
  linkcolor={blue},
  filecolor={Maroon},
  citecolor={blue},
  urlcolor={Blue},
  pdfcreator={LaTeX via pandoc}}

\author{}
\date{}

\begin{document}

\setstretch{1}
\title{Mechanism-Based Intelligence: Differentiable Incentives for Rational Coordination and Equilibrium Alignment in Multi-Agent Systems}

\author{
Stefano Grassi\textsuperscript{a}\thanks{Corresponding author: stefano.g@bu.ac.th}
}

\date{
\textsuperscript{a}Bangkok University, Phahonyothin Rd, Khlong Nueng, Khlong Luang District, Pathum Thani 12120, Thailand\\[2ex]
November 25, 2026
}

\maketitle

\begin{abstract}
Autonomous multi-agent systems struggle with the Hayekian Information problem (using dispersed private knowledge) and the Hurwiczian Incentive problem (aligning local actions with global objectives). I introduce Mechanism-Based Intelligence (MBI), an architecture that treats intelligence as emergent from the coordination of specialized modules rather than a single monolithic model. At its core, the Differentiable Price Mechanism computes the exact gradient of the system loss as a price; subtracting its own marginal cost, each agent obtains the incentive signal 

$$\mathbf{G}_i = - \frac{\partial \mathcal{L}_{\text{global}}}{\partial \mathbf{x}_i},$$ 

and the transfers take the Groves form. Each agent’s best response minimizes the global loss given the actions of others, so the negative global loss is an exact potential for the resulting game, with a unique equilibrium at the global optimum. Gradient-based play converges to it under a step-size condition when effort is costless, and stops near it otherwise: a satisficing rule halts each agent’s search when its marginal gain no longer exceeds its marginal computational cost. The equilibrium requires no knowledge of agents’ private costs, so asymmetric information causes no welfare loss; simultaneous gradient dynamics require only a public bound on cost curvature. Each iteration scales linearly ($\mathcal{O}(N)$) with the number of agents, though for losses summed over agents the iteration count also grows with $N$. For known, differentiable losses, MBI avoids the combinatorial search of Decentralized Partially Observable Markov Decision Processes. Under convexity and smoothness conditions, MBI aligns agent self-interest with the Planner’s objective, offering an auditable approach to multi-agent coordination.
\end{abstract}

 \vspace{1em}

\section{Introduction}\label{introduction}

Intelligence is a long-debated concept that traces back to Aristotle.
Since the inception of Artificial Intelligence (AI) in the 1950s,
definitions have ranged from problem-solving and goal achievement to
learning from data and adapting behavior to new environments. Broadly,
these perspectives fall into two camps: one seeks to replicate human
cognition (the connectionist camp), while the other aims to construct a
more abstract, rational form of intelligence.

The dominance of Deep Learning (DL) has reinforced the former, framing
intelligence primarily around individual cognition and single-agent
scaling. This single-agent perspective has carried over into multi-agent
systems, which are frequently treated as extensions of single-model
reasoning, sequential prediction, or centralized planning. Large
Language Models (LLMs) typify this approach: even when dealing with
interacting agents, they rely on unstructured natural-language exchange,
in which agents echo or adapt to each other's text without binding
commitments. Because these models operate as black-box predictors
reliant on pattern correlation, often substituting memorization for
causal understanding \citep{marcus2018deep}, token-level dialogue
provides no formal guarantee of strategic alignment or global
optimality.

Alternative cognitive architectures, such as LeCun's Hierarchical Joint
Embedding Predictive Architecture (H-JEPA) \citep{lecun2022path}, aim to
build centralized ``World Models'' that learn compact representations of
the environment. Yet these frameworks assume that a single unified model
can acquire and process all relevant information. Attempting to solve a
decentralized information-aggregation problem through centralization
creates two problems:

\begin{enumerate}
\def\labelenumi{\arabic{enumi}.}
\tightlist
\item
  The Hayekian Information Problem \citep{hayek1945use}: Centralized
  models lack access to the dispersed, tacit, and context-dependent
  knowledge held by decentralized actors (humans, sensors, sub-agents).
  As Polanyi notes, ``we know more than we can tell''
  \citep{polanyi1966tacit}; this tacit knowledge is required for optimal
  local action but cannot be captured purely through background
  observation.
\item
  The Hurwiczian Incentive Problem \citep{hurwicz1972centralized}:
  Cognitive architectures lack mechanisms to keep autonomous agents,
  acting on local knowledge, aligned with global goals. Classical
  mechanism design shows that strategic agents can manipulate collective
  decisions \citep{gibbard1973manipulation, satterthwaite1975strategy},
  and that efficiency, incentive compatibility (IC), voluntary
  participation, and budget balance cannot generally coexist
  \citep{green1977characterization, myerson1983efficient}. Achieving
  unmanipulable global outcomes requires building IC directly into the
  mechanism; in quasilinear settings, this is achieved via Groves
  transfers underlying Vickrey--Clarke--Groves (VCG) mechanisms
  \citep{vickrey1961counterspeculation, clarke1971multipart, groves1973incentives}.
  Without these constraints, agents prioritize local savings (such as
  minimizing compute) over system-wide performance, causing coordination
  and reliability failures.
\end{enumerate}

Real-world institutions face these same constraints. Countries and
organizations function as distributed systems of autonomous agents with
conflicting preferences, private information, and local limits. In these
settings, effective decision-making arises not from a central brain, but
from coordination and incentives, the exact dynamics connectionist
models try to replicate inside a single network. This requires shifting
the core question of AI from ``Can machines think?'' to ``How can
machines act coherently in structured, multi-agent environments toward
shared objectives?''

To address this, I propose Mechanism-Based Intelligence (MBI), which
frames intelligence as a product of coordination among autonomous
processes rather than individual cognition. MBI combines three economic
foundations: Hurwiczian mechanism design for goal alignment
\citep{hurwicz2006designing}, von Neumann--Morgenstern (VNM) utility
theory for rational choice under uncertainty \citep{von1944theory}, and
Simon's satisficing for bounded rationality (BR)
\citep{simon1947administrative, simon1955behavioral}. In MBI, a Planner
defines the system objective as a Differentiable Directed Acyclic Graph
(D-DAG), while computationally bounded agents maximize local utility. A
gradient-based feedback loop, the Differentiable Price Mechanism (DPM),
computes prices under which, given the conditions of Appendix B, the
unique equilibrium of the system is the global optimum. These guarantees
hold when the Planner knows the system loss, agents hold private costs,
and actions are real-valued and enter a differentiable, convex global
loss.

\section{The Cognitive-Economic
Divide}\label{sec-cognitive-economic-divide}

\subsection{The Cognitive Paradigm and Its
Limitations}\label{the-cognitive-paradigm-and-its-limitations}

The cognitive paradigm aims to emulate human thought through
comprehensive knowledge acquisition, tracing back to the neural networks
of McCulloch and Pitts \citeyearpar{mcculloch1943logical} and expanding
with modern compute and Machine Learning (ML). However, even within AI,
the assumption of a single centralized ``brain'' has drawn criticism.
Hiroaki Kitano's work on the RoboCup challenge
\citeyearpar{kitano1998robocup}, for instance, argues that real-world
scale requires massive parallelism, redundancy, and decentralized
self-organization \citep{kitano1994massively}, exposing the limits of
centralized aggregation in multi-agent settings.

\subsection{The Fragility of
Autoregression}\label{the-fragility-of-autoregression}

Autoregressive LLMs are central to this approach. These architectures
predict each token from previous ones, learning a conditional
distribution over the next token given the preceding context. While they
exhibit emergent abilities such as complex pattern matching, their
autoregressive nature poses serious challenges for long-horizon planning
and multi-step decision-making \citep{berti2025emergent}. Local reward
optimization, as in reward-based fine-tuning, can lead to undesirable
behaviors like deceptive planning, manipulative strategies or reward
hacking, an instance of the Hurwiczian Incentive Problem.

Unlike a Kalman filter, which updates its state estimate optimally for
linear-Gaussian state-space models, the LLM has no global correction
mechanism. Token generation is a local, step-by-step optimization of
coherence. Over long sequences, these local approximations produce an
approximation drift.

Assuming uniform, independent token errors, the probability of
maintaining full coherence across \(n\) tokens decays exponentially:
\(P(\text{coherence}) = (1 - e)^n\). Error compounding across multi-step
reasoning has been documented empirically and analyzed theoretically for
compositional tasks \citep{dziri2023faith}. Some research relaxes this
assumption, suggesting that errors concentrate at sparse ``key tokens''
representing decision points \citep{arbuzov2025beyond}. Concentration of
failure at these steps amplifies the risk. Empirical studies on Large
Reasoning Models (LRMs) describe this instability, showing declining
accuracy beyond certain compositional complexities and inconsistent
reasoning, suggesting that autoregressive modeling is limited for
long-horizon decisions \citep{shojaee2025illusion}.

\subsection{The Burden of World
Simulation}\label{the-burden-of-world-simulation}

To mitigate sequential errors, more sophisticated architectures
construct centralized, predictive ``World Models''. LeCun's H-JEPA is a
leading example, learning abstract, hierarchical, multi-modal
representations to manage environmental uncertainty
\citep{lecun2022path}. This model assumes that a single system can
acquire all knowledge needed to simulate the environment.

This centralization introduces key limitations in multi-agent
coordination and reliability:

\begin{itemize}
\tightlist
\item
  Representational Approximation and Implicit Misalignment: H-JEPA
  optimizes a latent representation \(z\) to minimize internal energy,
  approximating rather than accessing the true environment state. This
  process can favor coherence over external validity. When the system
  scales, local optimization drifts from global goals, reproducing the
  Hayekian and Hurwiczian problems.
\item
  Planning Constraints and BR: Even with an accurate model, finding
  optimal actions in a high-dimensional space is computationally
  expensive \citep{finn2017deep}. Gradient-based optimization through
  learned dynamics often fails, forcing reliance on search methods like
  Monte-Carlo Tree Search (MCTS), sampling planners, or dynamic
  programming (DP). Consistent with Herbert Simon's concept of BR
  \citep{simon1957models}, agents cannot compute exact optima and must
  settle for satisficing solutions.
\end{itemize}

\subsection{Decentralization and Price
Coordination}\label{decentralization-and-price-coordination}

The failure of centralized planners, seen in the General Problem Solver
(GPS) \citep{newell1959gps}, mirrors the classic economic problem of
resource allocation. Rather than scaling computation inside a single
agent, classical market theorists argue for solving this through
decentralization \citep{smith1776wealth, hayek1945use}. In markets,
prices coordinate decisions without requiring any participant to hold
complete global knowledge, while keeping optimization local and
tractable under BR. More generally, a \emph{Mechanism} can
operationalize this principle by establishing rules and incentives that
align self-interested local actions with global objectives.

Applied to AI, mechanism design may counteract approximation drift in
cognitive models via external incentives. In MBI, specialized BR agents,
such as LLM planners, symbolic math checkers, or code interpreters,
operate under a Planner that governs interactions and resource
allocation. When agent outputs feed into a differentiable global loss,
this structure can supply the global correction missing in standalone
autoregressive models.

\subsection{Mechanisms and Differentiable
Incentives}\label{mechanisms-and-differentiable-incentives}

Prior work in Distributed AI (DAI) and Multi-Agent Systems (MAS) lacks
formal guarantees of reliable coordination
\citep{stone2000multiagent, weiss2000multiagent}:

\begin{itemize}
\tightlist
\item
  Protocol-Based Coordination: Predefined protocols (e.g., Contract Net)
  break down when unexpected states arise or conflict with real-time
  local data.
\item
  Approximate Learning and Trial-and-Error Alignment: ML and
  Reinforcement Learning (RL) can approximate collective behavior, but
  without formal IC guarantees, systems remain vulnerable to local
  optimization drift, a limitation shared by the ``Reward is Enough''
  hypothesis \citep{silver2021reward} and similar reward-maximizing
  frameworks \citep{levine2021understanding}.
\end{itemize}

Economic theory offers an alternative by integrating incentives into
agent optimization, an idea developed in implementation theory
\citep{maskin2002implementation}, Automated Mechanism Design (AMD)
\citep{conitzer2003automated}, and Differentiable Economics (DE)
\citep{duetting2022optimal}. Unlike these methods, which compute or
learn a mechanism, MBI fixes a Groves-form transfer and derives the
incentive signal (\(\mathbf{G}_i\)) from the differentiable loss
gradient. This establishes IC: each agent's best response minimizes the
global loss given the actions of others (Theorem 1), improving
coordination without trial-and-error exploration.

MBI's equilibrium prices are marginal-externality (Pigouvian) prices
\citep{pigou1920economics}, and its dynamics belong to the family of
gradient-based price processes that includes the Arrow--Hurwicz--Uzawa
method \citep{arrow1958studies}, network utility maximization
\citep{kelly1998rate}, and dual decomposition
\citep{palomar2006tutorial}. The induced game is a potential game in the
sense of Monderer and Shapley \citeyearpar{monderer1996potential}. MBI
does not claim a new equilibrium concept. Instead, its contributions are
threefold: computing these prices with backpropagation across arbitrary
computation graphs of agents (Appendix C.12); integrating a
bounded-rationality stopping rule (Theorems 2 and 4); and applying this
framework to multi-agent coordination.

MBI is closest in spirit to economies of learning agents. Baum's Hayek
machine \citep{baum1999toward} coordinates agents that buy the right to
act in a market with conservation of money, while Chang et al.
\citep{chang2020decentralized} design a society of agents that trade
state-operation rights via Vickrey auctions so the global optimum
coincides with a Nash equilibrium. Both learn from scalar rewards in
sequential environments with unknown dynamics. MBI reaches the same
equilibrium alignment in a narrower setting, a known, differentiable
system loss over continuous actions, computing marginal-externality
prices exactly by backpropagation rather than discovering them through
bidding, while keeping agent costs private. MBI requires the convexity
and smoothness conditions of Appendix B and gives up budget balance
(Section B.3).

\section{Core Assumptions and Economic
Foundations}\label{core-assumptions-and-economic-foundations}

MBI rests on three economic foundations, which define how information,
incentives and computation are managed to align local actions with
globally optimal outcomes.

\subsection{Hurwiczian Mechanism
Design}\label{hurwiczian-mechanism-design}

MBI addresses the Hurwiczian Incentive Problem, the failure of locally
optimizing agents to align with global goals, using Mechanism Design
principles \citep{hurwicz2006designing}.

Key components:

\begin{itemize}
\tightlist
\item
  Institutional Role: The Planner defines the system loss
  (\(\mathcal{L}_{\text{system}}\)) and structures agent interactions
  via the D-DAG (Appendix B). Its task is not prediction, but mechanism
  design.
\item
  IC: The mechanism ensures that an agent's self-interested action,
  maximizing local utility, minimizes the global loss given the actions
  of the other agents, so that the unique equilibrium achieves the
  desired global outcome (Theorem 1).
\item
  Cost Assumption (Hurwiczian Constraint): Agents hold private
  operational knowledge (\(w_{i}\)), including their own costs, which
  remain hidden from the Planner due to tacit information. The Planner
  prices only the system loss, and each agent subtracts its own marginal
  cost, so the Planner requires no knowledge of agents' marginal costs
  (\(\lambda_i\)) and asymmetric information causes no welfare loss
  (Theorem 3). Simultaneous gradient dynamics additionally need a public
  bound on the curvature of these costs (Theorem 4). Where the Planner
  must instead price an agent's cost itself, basing it on the expected
  cost (\(E[\lambda_i]\)) under a probability distribution over agent
  types minimizes the expected global loss for costs affine in
  \(\lambda_i\) (Theorem 3; Appendix C.6).
\item
  The DPM: The mechanism derives the incentive signals, assuming convex
  action spaces, differentiable costs and a jointly strictly convex
  global loss. As a result, local actions (\(\mathbf{x}_i\)) align with
  \(\mathcal{L}_{\text{global}}\) while keeping agent costs private,
  allowing the system to function as a decentralized market for
  computational coordination. Throughout, alignment means alignment with
  the Planner's objective \(\mathcal{L}_{\text{global}}\); whether that
  objective reflects human values, the central question of AI alignment,
  is outside the scope of this paper.
\end{itemize}

\subsection{Von Neumann--Morgenstern Utility}\label{sec-vnm-utility}

MBI bypasses the intractable global predictions of cognitive models
using VNM Expected Utility Theory, formally modeling rational action
under uncertainty \citep{von1944theory}.

\begin{itemize}
\tightlist
\item
  The Divergence: Instead of predicting every possible future, MBI uses
  local, private information (\(w_{i}\)). Each agent's action
  (\(\mathbf{x}_i\)) is treated as a lottery, a probability distribution
  over outcomes given its best local knowledge.
\item
  Rational Choice: The agent selects \(\mathbf{x}_i\) to maximize
  expected utility (\(E[U_i]\)). This provides a tractable model for
  uncertainty, enabling efficient utilization of private information
  without eliciting it, thereby addressing the Hayekian Information
  Problem with respect to agents' private costs. In the formal analysis
  of Appendix B, actions and outcomes are deterministic, so expected
  utility reduces to the realized quasilinear utility \(U_i\);
  uncertainty enters only through the Planner's prior over private costs
  (Theorem 3) and, empirically, through persistent action noise
  (Appendix C.7). Extending the guarantees to agents whose actions are
  genuine lotteries over outcomes is left to later work.
\end{itemize}

\subsection{Simon's Satisficing}\label{simons-satisficing}

MBI incorporates BR \citep{simon1947administrative, simon1955behavioral}
to address the intractable search problem.

\begin{itemize}
\tightlist
\item
  Principle: Agents cannot compute the true optimum due to resource
  constraints; they stop searching once an action meets a sufficiency
  criterion.
\item
  MBI Innovation: Computational cost (\(\kappa(T)\)) is explicitly
  included in the agent's utility function, modeled as a differentiable
  function of planning effort \(T\) (e.g., iterations or search time).
  With \(U_i = t_i - C_i\) denoting the agent's utility from the scalar
  transfer \(t_i\) issued by the DPM (Section~\ref{sec-dpm}), the
  optimization becomes a continuous trade-off:
\end{itemize}

\[U(T) = t_i(\mathbf{x}_i(T), \mathbf{X}_{-i}) - C_i(\mathbf{x}_i(T)) - \kappa(T) \tag{1}\]

The agent stops searching once the marginal gain from further planning
no longer exceeds its marginal cost. Under gradient-ascent search and
convex \(\kappa(T)\), this stopping rule is optimal for the agent given
the other agents' actions (Theorem 2) and bounds the search effort each
agent spends, avoiding the architectural intractability of unbounded
search methods like MCTS or DP. This bounded-search behavior is
empirically validated in Appendix C.9.

\section{The MBI Architecture}\label{the-mbi-architecture}

\subsection{The Differentiable Directed Acyclic Graph
Structure}\label{the-differentiable-directed-acyclic-graph-structure}

MBI is represented as a D-DAG. This structure enforces causality and
differentiability across the network:

\begin{itemize}
\tightlist
\item
  Nodes (\(\mathcal{N}\)): Each node represents a distinct computational
  unit, which can be an external data source, a deterministic function,
  or an autonomous agent (\(A_i\)).
\item
  Edges (\(\mathcal{E}\)): Directed edges define the flow of actions
  (\(\mathbf{x}_i\)) and information between agents. The directional,
  acyclic nature ensures a well-defined computational sequence and
  enables the application of the chain rule.
\end{itemize}

\subsection{The Planner and Global
Loss}\label{the-planner-and-global-loss}

MBI is initiated by the Planner (\(P\)), whose role, according to
Hurwiczian principles, is deliberately minimal. The Planner's function
is restricted to:

\begin{enumerate}
\def\labelenumi{\arabic{enumi}.}
\tightlist
\item
  Defining the structure (D-DAG).
\item
  Defining the system loss (\(\mathcal{L}_{\text{system}}\)): Calculated
  at the terminal node(s) of the D-DAG and represents the system
  objective. Together with the agents' private costs
  \(C_j(\mathbf{x}_j)\), known only to each agent, it defines the global
  loss \(\mathcal{L}_{\text{global}}\), which represents the overall
  objective (Eq. (2)). Because agents have no incentive to disclose
  \(C_{j}\) truthfully and the Planner lacks authority to dictate
  actions, \(\mathcal{L}_{\text{global}}\) cannot be minimized
  centrally; instead, the Planner prices only
  \(\mathcal{L}_{\text{system}}\), while each agent accounts for its own
  marginal cost to reach the global optimum (Theorems 1 and 3). When all
  costs are public and actions can be centrally assigned, MBI reduces to
  distributed computation rather than providing an informational
  advantage.
\end{enumerate}

\subsection{The Differentiable Price Mechanism}\label{sec-dpm}

The DPM is the core mechanism that translates economic principles into
continuous computation via a two-stage process: the Forward Pass and the
Backward Pass.

\subsubsection{Forward Pass: Action and Outcome
Generation}\label{forward-pass-action-and-outcome-generation}

In the Forward Pass, agents select actions under the mechanism's
structure, which collectively determine the realized outcome or global
loss.

The process begins when an agent \(A_i\) receives its current inputs
(from preceding nodes/agents). Based on these inputs and its private
information (\(w_i\)), the agent computes its action \(\mathbf{x}_i\) by
maximizing its expected utility \(E[U_i]\) (as defined in
Section~\ref{sec-vnm-utility}). This action \(\mathbf{x}_i\) is then
passed forward through the D-DAG, culminating in the system loss
\(\mathcal{L}_{\text{system}}(\mathbf{X})\), which combines with the
agents' private costs to form the global loss:

\[\mathcal{L}_{\text{global}}(\mathbf{X}) = \mathcal{L}_{\text{system}}(\mathbf{X}) + \sum_{j=1}^{N} C_j(\mathbf{x}_j) \tag{2}\]

\subsubsection{Backward Pass: Computation of the Incentive
Signal}\label{backward-pass-computation-of-the-incentive-signal}

The Backward Pass uses the chain rule to compute the incentive signal
\(\mathbf{G}_i\), propagating the marginal impact of each agent's action
on the system loss back through the D-DAG (Appendix C.12).

These incentive signals are not rewards in the usual sense, nor are they
used to update network weights as in standard DL. They encode each
agent's marginal contribution to the global objective, so that
self-interested actions align with the Planner's goals: the DPM turns a
global optimization problem into local guidance.

Prices are partial derivatives with the other agents' actions held
fixed: edges of the D-DAG describe how actions combine in the system
loss, and the Backward Pass differentiates these known system maps, not
the agents' decision rules, so no agent's cost or reaction function
enters any price and agents can update their actions simultaneously.

\subsubsection{The Gradient as the Incentive
Signal}\label{the-gradient-as-the-incentive-signal}

The incentive signal \(\mathbf{G}_i\) is mathematically defined as the
negative gradient of the global loss with respect to the agent's output
action \(\mathbf{x}_i\):

\[\mathbf{G}_i = - \frac{\partial \mathcal{L}_{\text{global}}}{\partial \mathbf{x}_i} \tag{3}\]

Since no other agent's cost depends on \(\mathbf{x}_i\),
\(\mathbf{G}_i\) is the sum of the Planner's price
\(-\frac{\partial \mathcal{L}_{\text{system}}}{\partial \mathbf{x}_i}\),
computed by the Backward Pass, and the negative of the agent's own
marginal cost \(-\nabla_{\mathbf{x}_i} C_i\), which the agent adds
locally. The price is the gradient of the scalar transfer
\(t_i(\mathbf{x}_i, \mathbf{X}_{-i}) = -\mathcal{L}_{\text{system}}(\mathbf{x}_i, \mathbf{X}_{-i}) + h_i(\mathbf{X}_{-i})\),
where \(h_i\) does not depend on \(\mathbf{x}_i\); the agent's utility
is \(U_i = t_i - C_i(\mathbf{x}_i)\), and
\(\mathbf{G}_i = \nabla_{\mathbf{x}_i} U_i\) is its net marginal utility
(Appendix B.1.1).

\begin{itemize}
\tightlist
\item
  Economic Interpretation of Internalizing Externalities: The term
  \(\frac{\partial \mathcal{L}_{\text{system}}}{\partial \mathbf{x}_i}\)
  represents the marginal externality, the cost (or benefit) imposed on
  the system objective by a slight change in the agent's local action
  \(\mathbf{x}_i\). By paying the agent the negative of this cost, the
  DPM provides a signal that measures the global value the agent
  contributes toward minimizing the global loss, so that the agent
  internalizes all externalities its action creates.
\end{itemize}

\paragraph{Hurwiczian Incentive
Compatibility}\label{hurwiczian-incentive-compatibility}

Under the conditions of Appendix B, the continuous, gradient-based
signal delivers Hurwiczian IC by aligning the agent's local incentives
with the global goal.

Since the agent chooses \(\mathbf{x}_i\) to maximize its utility
\(U_i = t_i - C_i\), and since \(U_i\) and
\(-\mathcal{L}_{\text{global}}\) differ only by terms that do not depend
on \(\mathbf{x}_i\):

\[\underset{\mathbf{x}_i}{\operatorname{argmax}}\, U_i(\mathbf{x}_i, \mathbf{X}_{-i}) = \underset{\mathbf{x}_i}{\operatorname{argmin}}\, \mathcal{L}_{\text{global}}(\mathbf{x}_i, \mathbf{X}_{-i}) \quad \text{for every } \mathbf{X}_{-i} \tag{4}\]

Each gradient step on \(U_i\) is a step of steepest descent on
\(\mathcal{L}_{\text{global}}\) in \(\mathbf{x}_i\). Under the
conditions of Appendix B, \(-\mathcal{L}_{\text{global}}\) serves as an
exact potential function, so the unique equilibrium coincides with the
global minimum of \(\mathcal{L}_{\text{global}}\) (Theorem 1).

\subsubsection{The High-Information
Feedback}\label{the-high-information-feedback}

The incentive signal \(\mathbf{G}_i\) is a vector (or tensor) of the
same dimension as the action \(\mathbf{x}_i\), specifying the direction
and magnitude of the steepest local change in every component to reduce
the global loss. When the global loss is known and differentiable, this
is far richer and more actionable than the low-information, scalar
reward signal typically used in standard RL.

\subsection{The MBI Optimization
Cycle}\label{the-mbi-optimization-cycle}

\begin{figure}[H]

\centering{

\includegraphics[width=0.5\linewidth,height=\textheight,keepaspectratio]{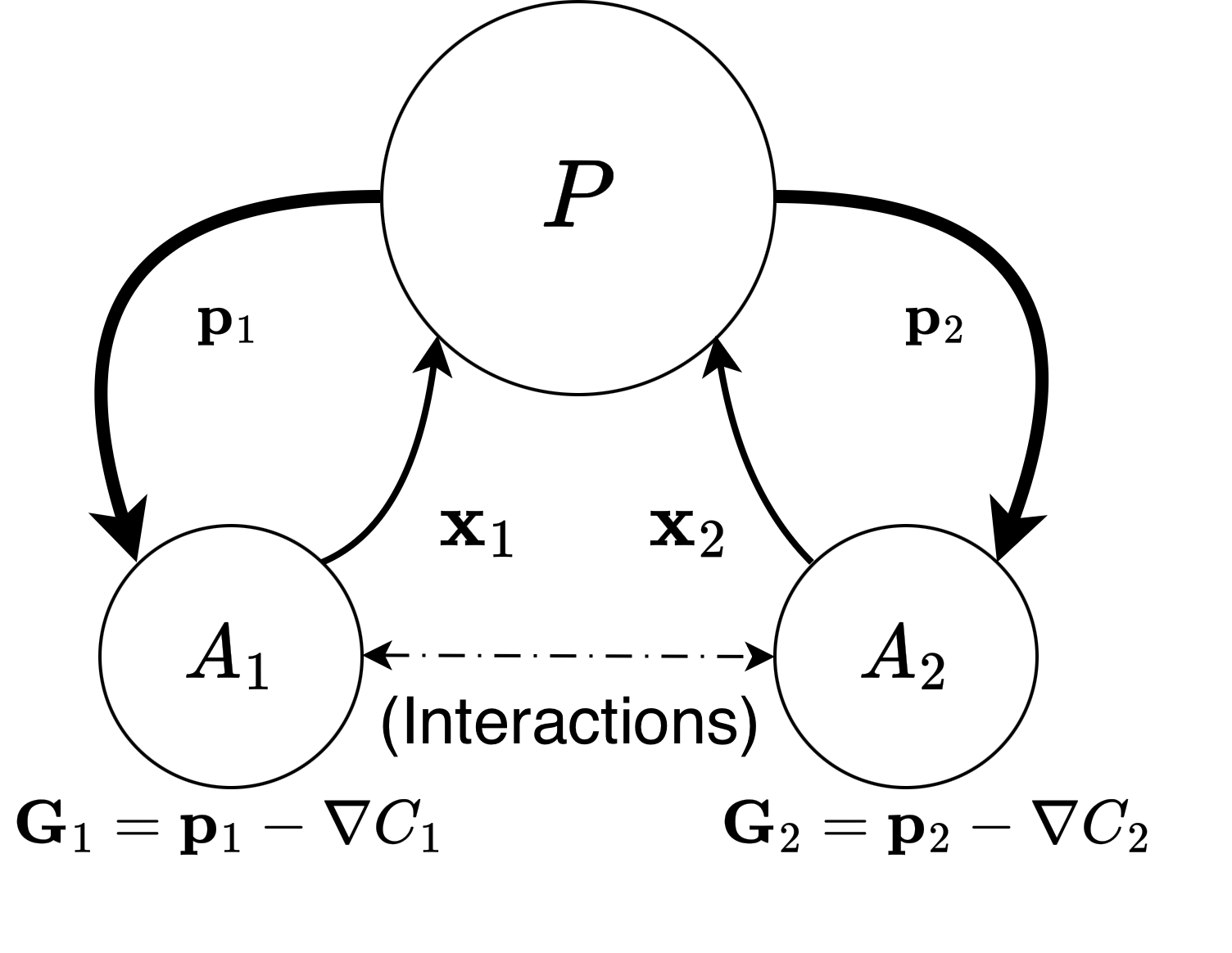}

}

\caption{\label{fig-mbi}Atomic MBI Architecture and DPM. This diagram
illustrates the minimal, closed-loop interaction between the Planner
(\(P\)) and two Agents (\(A_1, A_2\)). In the Forward Pass (thin
arrows), the agents' actions \(\mathbf{x}_1\) and \(\mathbf{x}_2\) reach
the Planner. The dotted two-way arrow marks the strategic interaction
between the agents: through the system loss, each agent's incentive
depends on the other's action. In the Backward Pass (bold arrows), the
DPM delivers each agent's price
\(\mathbf{p}_i = -\nabla_{\mathbf{x}_i}\mathcal{L}_{\text{system}}\);
each agent subtracts its own marginal cost to form its incentive signal
\(\mathbf{G}_i = \mathbf{p}_i - \nabla_{\mathbf{x}_i} C_i = -\nabla_{\mathbf{x}_i}\mathcal{L}_{\text{global}}\),
so the Planner never needs the agents' costs.}

\end{figure}

MBI runs as a closed loop, illustrated in Figure~\ref{fig-mbi}, that
replaces the traditional, iterative planning cycle:

\begin{enumerate}
\def\labelenumi{\arabic{enumi}.}
\tightlist
\item
  Planner Configures: Defines the D-DAG and
  \(\mathcal{L}_{\text{system}}\).
\item
  Agent Action (Forward Pass): Each agent \(A_i\) updates its
  utility-maximizing action \(\mathbf{x}_i\), constrained by the
  Satisficing Condition (BR), using the price received in the previous
  cycle and its own marginal cost. The agent keeps searching while the
  marginal incentive gain exceeds the marginal cost of effort, and halts
  when \(\frac{\partial U}{\partial T} \leq 0\), where
  \(U(T) = U_i - \kappa(T)\) is its utility net of effort cost (Theorem
  2). During adjustment, agents respond to current prices rather than
  anticipating the effect of their actions on future prices; strategic
  manipulation of the adjustment path is not analyzed.
\item
  The DPM computes (Backward Pass):
  \[-\frac{\partial \mathcal{L}_{\text{system}}}{\partial \mathbf{x}_i}, \qquad \mathbf{G}_i = -\frac{\partial \mathcal{L}_{\text{system}}}{\partial \mathbf{x}_i} - \nabla_{\mathbf{x}_i} C_i = - \frac{\partial \mathcal{L}_{\text{global}}}{\partial \mathbf{x}_i} \tag{5}\]
  This calculation sets the price equal to the negative marginal
  externality, so that, under the conditions of Theorem 1, the unique
  equilibrium of the induced game is the global optimum.
\item
  Incentive Delivery: The computed price is delivered to the agent,
  updating its expected utility for the next cycle.
\item
  System Update: The system state is updated, and the cycle repeats,
  moving the collective action toward the minimum of
  \(\mathcal{L}_{\text{global}}\) until it comes to rest (Theorem 4).
\end{enumerate}

\subsection{Computational Complexity and
Scaling}\label{computational-complexity-and-scaling}

The MBI architecture offers computational advantages over alternative
methods. Because the D-DAG ensures that the DPM only requires local
computation of the gradient at each agent, the cost of each iteration
scales almost linearly with the size of the network, (\(\mathcal{O}(N)\)
when each agent has a bounded number of connections. The number of
iterations depends on the step size, which for losses summed over agents
shrinks as \(1/N\) (Theorem 4). Where the global loss is known and
differentiable, MBI avoids the combinatorial search over joint policies
of Dec-POMDPs \citep{bernstein2002}, which address a different setting
with stochastic, partially observable transitions. Empirically, the cost
of one iteration grows close to linearly with \(N\) up to \(10^7\)
agents, while for summed losses the number of iterations to converge
grows roughly in proportion to \(N\) (Appendix C.11). Linear cost per
iteration is essential for applying MBI to large-scale multi-agent
coordination problems.

\section{Toy Example: The Two-Agent Assembly
Line}\label{toy-example-the-two-agent-assembly-line}

To ground the abstract mechanism design principles, the operational
framework is illustrated using a simplified, sequential two-agent
manufacturing task. This example demonstrates how MBI replaces
centralized planning with a provably aligned incentive structure
computed through differentiation.

\subsection{System Setup and Planner
Objective}\label{system-setup-and-planner-objective}

The Planner (\(P\)) oversees a sequential assembly process involving two
specialized agents, \(A_1\) and \(A_2\). The system's architecture is a
simple D-DAG: \(A_1\) produces an intermediate output \(\mathbf{x}_1\),
which serves as input for \(A_2\), which then produces the final output
used to calculate the loss.

\begin{itemize}
\tightlist
\item
  \(A_1\) (Planning Agent): Chooses the first component size
  \(\mathbf{x}_1 \in \mathbb{R}\).
\item
  \(A_2\) (Refinement Agent): Chooses the final adjustment
  \(\mathbf{x}_2 \in \mathbb{R}\).
\end{itemize}

The objective is to minimize the combined cost of error from the target
output (\(Y^*\)), defined by \(P\) as the system loss
\(\mathcal{L}_{\text{system}} = (\mathbf{x}_1 + \mathbf{x}_2 - Y^*)^2\),
and the inherent inefficiency introduced by \(\mathbf{x}_1\), which is
\(A_1\)'s private cost \(C_1(\mathbf{x}_1) = \lambda \mathbf{x}_1^2\)
(\(A_2\) has no cost, \(C_2 = 0\)):

\[\mathcal{L}_{\text{global}}(\mathbf{x}_1, \mathbf{x}_2) = (\mathbf{x}_1 + \mathbf{x}_2 - Y^*)^2 + \lambda \mathbf{x}_1^2 \tag{6}\]

Here, \(Y^*\) is the target value and \(\lambda > 0\) is a penalty
coefficient (e.g., resource cost of a large component), known only to
\(A_1\). Each rational agent chooses \(\mathbf{x}_i\) to maximize its
utility \(U_i = t_i - C_i\) net of computational effort cost
\(\kappa(T)\) (Eq. (1)).

\subsection{Incentive Computation: The Backward Pass as
Pricing}\label{incentive-computation-the-backward-pass-as-pricing}

\(P\) computes the price for each agent via the DPM across the
differentiable graph; each agent subtracts its own marginal cost to
obtain its incentive signal \(\mathbf{G}_i\).

\subsubsection{\texorpdfstring{Agent \(A_2\)'s Direct Incentive
(\(\mathbf{G}_2\))}{Agent A\_2's Direct Incentive (\textbackslash mathbf\{G\}\_2)}}\label{agent-a_2s-direct-incentive-mathbfg_2}

\(A_2\) is the final agent. Since \(C_2 = 0\), its incentive is \(P\)'s
price, the negative gradient of the loss with respect to its action
\(\mathbf{x}_2\):

\[\mathbf{G}_2 = - \frac{\partial \mathcal{L}_{\text{global}}}{\partial \mathbf{x}_2} = - 2(\mathbf{x}_1 + \mathbf{x}_2 - Y^*) \tag{7}\]

\emph{Economic Interpretation: \(A_2\) receives an incentive signal
based solely on its marginal contribution to closing the gap between
total output and the target \(Y^*\).}

\subsubsection{\texorpdfstring{Agent \(A_1\)'s Total Incentive
(\(\mathbf{G}_1\)): Internalizing the
Externality}{Agent A\_1's Total Incentive (\textbackslash mathbf\{G\}\_1): Internalizing the Externality}}\label{agent-a_1s-total-incentive-mathbfg_1-internalizing-the-externality}

\(A_1\)'s action \(\mathbf{x}_1\) impacts the loss both directly
(through the resource cost \(\lambda \mathbf{x}_1^2\)) and indirectly
(by altering the final error term). Applying the sum and chain rules:

\[\frac{\partial \mathcal{L}_{\text{global}}}{\partial \mathbf{x}_1} = \underbrace{2(\mathbf{x}_1 + \mathbf{x}_2 - Y^*) \cdot 1}_{\text{Marginal Externality (on Target Error)}} + \underbrace{2 \lambda \mathbf{x}_1}_{\text{Marginal Direct Cost (on Resource Penalty)}} \tag{8}\]

Therefore, \(A_1\)'s incentive signal is:

\[\mathbf{G}_1 = - \frac{\partial \mathcal{L}_{\text{global}}}{\partial \mathbf{x}_1} = - \left[ 2(\mathbf{x}_1 + \mathbf{x}_2 - Y^*) + 2 \lambda \mathbf{x}_1 \right] \tag{9}\]

\begin{quote}
Mechanism Insight: The first term,
\(2(\mathbf{x}_1 + \mathbf{x}_2 - Y^*)\), represents the externality
imposed by \(A_1\)'s choice on the final global outcome (which \(A_1\)
would ignore in an unincentivized regime). The DPM computes this
marginal impact and delivers it as \(A_1\)'s price, so that \(A_1\)
internalizes this global cost. The second term,
\(2 \lambda \mathbf{x}_1\), is \(A_1\)'s own marginal cost, which
\(A_1\) subtracts locally: \(P\) never needs to know \(\lambda\).
\end{quote}

\subsection{Equilibrium Solution and Bounded
Search}\label{equilibrium-solution-and-bounded-search}

Under the conditions of Theorem 4, when search effort is costless
(\(\frac{\partial \kappa}{\partial T} = 0\)), the agents collectively
drive the system to the unique equilibrium where both incentive signals
vanish (\(\mathbf{G}_1 = 0\) and \(\mathbf{G}_2 = 0\)):

\begin{enumerate}
\def\labelenumi{\arabic{enumi}.}
\item
  Setting \(\mathbf{G}_2 = 0\) (Local Alignment):
  \[-2(\mathbf{x}_1^* + \mathbf{x}_2^* - Y^*) = 0 \quad \implies \quad \mathbf{x}_1^* + \mathbf{x}_2^* = Y^* \tag{10}\]
\item
  Setting \(\mathbf{G}_1 = 0\) (Global Alignment via Internalization):
  Substituting \(\mathbf{x}_1^* + \mathbf{x}_2^* = Y^*\) into Eq. (9):
  \[- \left[ \underbrace{2(\mathbf{x}_1^* + \mathbf{x}_2^* - Y^*)}_{=0} + 2 \lambda \mathbf{x}_1^* \right] = 0 \quad \implies \quad 2 \lambda \mathbf{x}_1^* = 0 \tag{11}\]
  Since \(\lambda > 0\), this requires \(\mathbf{x}_1^* = 0\).
\end{enumerate}

The agents' incentives thus lead to the unique global optimum
\(\mathbf{x}_1^* = 0\) and \(\mathbf{x}_2^* = Y^*\): the costly
component \(\mathbf{x}_1\) is not used, and the zero-cost agent \(A_2\)
delivers the entire target \(Y^*\). When computational effort is costly
(\(\frac{\partial \kappa}{\partial T} > 0\)), each agent, searching
along \(\mathbf{G}_i\) at rate \(\eta\), instead halts refinement once
its marginal gain \(\eta \|\mathbf{G}_i\|^2\) no longer exceeds
\(\frac{\partial \kappa}{\partial T}\) (Theorem 2), stopping within the
residual bound
\(\|\mathbf{G}_i\| \le \sqrt{\frac{1}{\eta}\frac{\partial \kappa}{\partial T}}\)
rather than performing unbounded search (Appendix C.1).

\section{Conclusion}\label{conclusion}

As AI systems become increasingly multi-agent \citep{dafoe2020open},
purely cognitive architectures remain limited by dispersed private
knowledge (the Hayekian problem) and misaligned local incentives (the
Hurwiczian problem). MBI addresses a specific form of both problems by
framing multi-agent systems as a differentiable game rather than a
centralized prediction task.

\subsection{Empirical Validation}\label{empirical-validation}

The theory is supported by numerical experiments (Appendix C).

Key findings include:

\begin{itemize}
\tightlist
\item
  IC and BR Confirmation: Agents' actions converge to the global optimum
  when effort is costless, and come to rest within the residual bound of
  the Satisficing Condition when it is costly, fulfilling IC and BR
  requirements.
\item
  Computational Advantage: The cost of each iteration grows close to
  linearly (\(\mathcal{O}(N)\)) with the number of agents, while for
  losses summed over agents the number of iterations grows with \(N\)
  (Appendix C.11).
\item
  Robustness and Equivalence: Under persistent action noise the dynamics
  remain stable, with a stationary excess loss predicted in closed form,
  and a private, time-varying cost unknown to the Planner is
  internalized (Appendix C.7, C.10). The equilibrium coincides with the
  efficient outcome that VCG implements.
\item
  Private Costs: At equilibrium, the Planner requires no knowledge of
  agents' costs, and asymmetric information causes no welfare loss
  (Theorem 3); where the Planner must instead price an agent's cost
  itself, basing it on the expected cost minimizes the expected global
  loss when costs are affine in the private parameter.
\end{itemize}

\subsection{Potential Applications}\label{potential-applications}

MBI could apply to domains requiring coordinated optimization over
differentiable objectives, such as supply chain routing, automated
market making, decentralized autonomous organizations, and power grid
dispatch, wherever the global objective can be expressed as a
differentiable loss. Because every agent updates along an explicit
gradient signal \(\mathbf{G}_i\), the resulting decisions are
transparent and auditable, supporting goals in Explainable AI
\citep{gunning2017explainable} and Trustworthy AI
\citep{ali2023explainable}.

\subsection{Limitations and Future
Work}\label{limitations-and-future-work}

MBI's theoretical guarantees rely on assumptions that define its current
scope:

\begin{itemize}
\tightlist
\item
  Smoothness and Convexity: The DPM requires twice continuous
  differentiability (\(C^2\) smoothness) of
  \(\mathcal{L}_{\text{global}}\) and all cost functions
  \(C_i(\mathbf{x}_i)\), excluding discrete actions or discontinuous
  rewards. Global convergence (Section B.6) further requires joint
  strict convexity and a step size \(0 < \eta < 2/L_{\nabla}\); in
  non-convex settings, gradient dynamics may stall at local minima
  (Appendix C.8), and for losses summed over agents the step size scales
  as \(1/N\), increasing iteration counts with \(N\).
\item
  Equilibrium and Path-Manipulation Scope: Alignment is a Nash
  equilibrium property rather than dominant-strategy incentive
  compatibility, as agents choose continuous actions rather than report
  types (Section B.3). Transfers depend only on realized outcomes;
  strategic manipulation of the adjustment path, such as strategic early
  stopping, in which an agent inflates its apparent effort cost so that
  others adjust around it, is not analyzed.
\item
  Budget Balance, Participation, and Priors: Transfers are not budget
  balanced, and voluntary participation is not investigated (Section
  B.3). The Planner needs no prior when agents internalize costs locally
  (Section B.5); pricing an agent's cost directly still requires an
  accurate prior \(f(\lambda_i)\), which may fail under deep
  uncertainty.
\end{itemize}

Future work should relax these assumptions along four directions:

\begin{enumerate}
\def\labelenumi{\arabic{enumi}.}
\tightlist
\item
  Non-convex objectives: Extending global convergence guarantees to
  non-convex losses using trust-region or stochastic methods.
\item
  Type-reporting extensions: Adding a reporting stage to achieve
  dominant-strategy IC via Groves transfers or Bayesian IC via
  expected-externality transfers.
\item
  Higher-order beliefs: Incorporating dynamic Planner belief updates and
  multi-agent beliefs over private cost distributions.
\item
  Human-in-the-loop systems: Adapting the mechanism to hybrid human--AI
  teams using behavioral and organizational models of bounded search.
\end{enumerate}

\section{References}\label{references}

\renewcommand{\bibsection}{}
\bibliography{references.bib}

\section*{Appendix}\label{appendix}
\addcontentsline{toc}{section}{Appendix}

This Appendix provides the full technical documentation of MBI. Section
A summarizes the framework, Section B presents the formal derivations of
the paper's claims, Section C provides the empirical validation and
Section D addresses reproducibility.

\subsection*{Notation Convention}\label{notation-convention}
\addcontentsline{toc}{subsection}{Notation Convention}

Table 1 summarizes the notation used throughout the appendix. Bold
symbols denote vectors and non-bold symbols denote scalars.

\begin{table}[H]
    \centering
    \caption{Notation Convention}
    \begin{tabular}{|c|m{5.2cm}|m{5.5cm}|}
        \hline
        \textbf{Notation} & \textbf{Meaning} & \textbf{Context} \\
        \hline \hline
        $C_i(\mathbf{x}_i)$ & Denotes the economic cost of agent $A_i$'s action. & Used in $\mathcal{L}_{\text{global}}$ and in computing IC. \\
        \hline
        $\kappa(T)$ & Denotes the internal computational cost or effort. & Used in the Formal Satisficing Condition for BR, where $T$ is the effort level. \\
        \hline
        $t_i$, $h_i(\mathbf{X}_{-i})$ & Scalar transfer issued by the DPM, $t_i = -\mathcal{L}_{\text{system}} + h_i(\mathbf{X}_{-i})$, where $h_i$ does not depend on $\mathbf{x}_i$. & Agent utility $U_i = t_i - C_i$ (B.1.1, Theorem 1). \\
        \hline
        $\mathbf{G}_i$ & Incentive signal, $\mathbf{G}_i = -\nabla_{\mathbf{x}_i}\mathcal{L}_{\text{global}} = \nabla_{\mathbf{x}_i} U_i$: the agent's net marginal utility. & DPM; search direction in Theorems 2 and 4. \\
        \hline
        $\eta$ & Step size of the agents' gradient search. & Theorems 2 and 4. \\
        \hline
        $L_{\nabla}$ & Lipschitz constant of $\nabla \mathcal{L}_{\text{global}}$. & Step-size condition $0 < \eta < 2/L_{\nabla}$ (Theorem 4). \\
        \hline
    \end{tabular}
\end{table}

\subsection*{A: Mechanism-Based Intelligence
Summary}\label{a-mechanism-based-intelligence-summary}
\addcontentsline{toc}{subsection}{A: Mechanism-Based Intelligence
Summary}

Section A provides a consolidated overview of the key components and
operational logic of MBI for reference. The core elements of the
mechanism are listed in Table 2.

\subsubsection*{A.1 Core Mechanism}\label{a.1-core-mechanism}
\addcontentsline{toc}{subsubsection}{A.1 Core Mechanism}

\begin{table}[H]
    \centering
    \caption{Core Components of the MBI Mechanism}
    \begin{tabular}{|c|m{5.2cm}|m{5.5cm}|}
        \hline
        \textbf{Component} & \textbf{Description} & \textbf{Economic Principle} \\
        \hline \hline
        Planner ($P$) & Defines the system loss $\mathcal{L}_{\text{system}}$ (which, together with the agents' private costs, forms $\mathcal{L}_{\text{global}}$) and the Differentiable Directed Acyclic Graph (D-DAG) encoding informational and causal dependencies among agents. Its role is institutional only: it defines the mechanism without optimizing. & Hurwiczian IC \\
        \hline
        Agent ($A_i$) & A rational VNM actor maximizing local utility ($U_i$). It holds private knowledge ($w_i$), including its cost $C_i$. & VNM Expected Utility Theory \\
        \hline
        Utility Function ($U_i$) & $U_i = t_i - C_i(\mathbf{x}_i)$, quasilinear in the scalar transfer $t_i$. During search, the agent maximizes $U_i - \kappa(T)$, balancing utility gain against internal computational cost ($\kappa$). & Quasilinear utility (Groves); Simon's BR (Satisficing) \\
        \hline
        Incentive $\mathbf{G}_i$ & The agent's net marginal utility, $\mathbf{G}_i = -\nabla_{\mathbf{x}_i}\mathcal{L}_{\text{global}}$, formed from the Planner's price $-\nabla_{\mathbf{x}_i}\mathcal{L}_{\text{system}}$ minus its own marginal cost. & DPM \\
        \hline
    \end{tabular}
\end{table}

\subsubsection*{A.2 Algorithm}\label{a.2-algorithm}
\addcontentsline{toc}{subsubsection}{A.2 Algorithm}

\begin{algorithm}[H]
\caption{The MBI Optimization Cycle}
\begin{algorithmic}[1]
\Require System Loss Function $\mathcal{L}_{\text{system}}(\mathbf{X})$, D-DAG, step size $0 < \eta < 2/L_{\nabla}$, Agent Private Info $w$ (including $C_i$, known only to $A_i$)
\State Initialize agent actions $\mathbf{X}^{(0)}$; Planner computes initial prices $-\partial \mathcal{L}_{\text{system}} / \partial \mathbf{x}_i^{(0)}$
\While{some agent updated in the previous iteration} \Comment{Agent-Planner iterations until the system comes to rest (Theorem 4)}
    \State Forward Pass: for each agent $A_i$ (simultaneously)
    \State $\mathbf{G}_i^{(t)} = -\frac{\partial \mathcal{L}_{\text{system}}}{\partial \mathbf{x}_i^{(t)}} - \nabla_{\mathbf{x}_i} C_i(\mathbf{x}_i^{(t)})$ \Comment{Planner's price minus the agent's own marginal cost (B.5)}
    \If{$\eta \|\mathbf{G}_i^{(t)}\|^2 > \frac{\partial \kappa}{\partial T}$} \Comment{Satisficing Check (B.13)}
        \State $\mathbf{x}_i^{(t+1)} = \mathbf{x}_i^{(t)} + \eta\, \mathbf{G}_i^{(t)}$ \Comment{Gradient step on $U_i$}
    \Else
        \State $\mathbf{x}_i^{(t+1)} = \mathbf{x}_i^{(t)}$ \Comment{Agent satisfices}
    \EndIf
    \State Collective action $\mathbf{X}^{(t+1)} = (\mathbf{x}_1^{(t+1)}, \dots, \mathbf{x}_N^{(t+1)})$ is executed
    \State Compute System Loss: $\mathcal{L}_{\text{system}}^{\text{new}} = \mathcal{L}_{\text{system}}(\mathbf{X}^{(t+1)})$
    \State Backward Pass: for each agent $A_i$ in reverse topological order
    \State Planner computes the price $-\frac{\partial \mathcal{L}_{\text{system}}}{\partial \mathbf{x}_i^{(t+1)}}$ by backpropagation on the D-DAG and delivers it to $A_i$ \Comment{DPM signal: negative marginal externality of $A_i$'s action on the system loss}
\EndWhile
\Ensure Rest profile satisfying (B.14) for all agents; $\mathbf{X}^*$ if $\frac{\partial \kappa}{\partial T} = 0$
\end{algorithmic}
\end{algorithm}

\subsection*{B: Formal Derivations for MBI Incentive Compatibility,
Bounded Rationality, Private Costs and
Convergence}\label{b-formal-derivations-for-mbi-incentive-compatibility-bounded-rationality-private-costs-and-convergence}
\addcontentsline{toc}{subsection}{B: Formal Derivations for MBI
Incentive Compatibility, Bounded Rationality, Private Costs and
Convergence}

This section presents the formal mathematical justification for the core
claims of MBI: IC, BR, the treatment of private costs under asymmetric
information and convergence.

\subsubsection*{B.1 Regularity Conditions and Core
Assumptions}\label{b.1-regularity-conditions-and-core-assumptions}
\addcontentsline{toc}{subsubsection}{B.1 Regularity Conditions and Core
Assumptions}

\paragraph*{B.1.1 System and Agent Domain
Definitions}\label{b.1.1-system-and-agent-domain-definitions}
\addcontentsline{toc}{paragraph}{B.1.1 System and Agent Domain
Definitions}

\begin{itemize}
\tightlist
\item
  Action Space (\(\mathcal{X}\)): The agent's action \(\mathbf{x}_i\)
  lies in its action space \(\mathcal{X}_i \subset \mathbb{R}^k\), which
  is non-empty, convex, and either compact or equal to \(\mathbb{R}^k\).
  The joint action space is the product
  \(\mathcal{X} = \mathcal{X}_1 \times \dots \times \mathcal{X}_N\), so
  that any unilateral deviation by \(A_i\) remains feasible for every
  \(\mathbf{X}_{-i}\). The social optimum \(\mathbf{X}^*\) is assumed to
  lie in the interior of \(\mathcal{X}\); boundary optima are outside
  the scope of this appendix.
\item
  Agent Utility: Agents possess quasilinear utility in a scalar transfer
  \(t_i\) issued by the DPM:
  \(U_i(\mathbf{x}_i, \mathbf{X}_{-i}) = t_i(\mathbf{x}_i, \mathbf{X}_{-i}) - C_i(\mathbf{x}_i)\),
  with
  \(t_i(\mathbf{x}_i, \mathbf{X}_{-i}) = -\mathcal{L}_{\text{system}}(\mathbf{x}_i, \mathbf{X}_{-i}) + h_i(\mathbf{X}_{-i})\),
  where \(h_i\) does not depend on \(\mathbf{x}_i\). The incentive
  signal
  \(\mathbf{G}_i = -\nabla_{\mathbf{x}_i}\mathcal{L}_{\text{global}}\)
  is the agent's net marginal utility,
  \(\mathbf{G}_i = \nabla_{\mathbf{x}_i} U_i\), not its utility.
\item
  Information: Agent costs \(C_i(\mathbf{x}_i)\) are independent private
  values (IPV), known only to agent \(A_i\). Each \(C_i\) depends only
  on \(\mathbf{x}_i\) and on \(A_i\)'s private type, and
  \(\mathcal{L}_{\text{system}}\) depends on no private type. The
  Planner therefore computes only
  \(-\nabla_{\mathbf{x}_i}\mathcal{L}_{\text{system}}\), and agent
  \(A_i\) subtracts its own marginal cost \(\nabla_{\mathbf{x}_i} C_i\)
  locally.
\end{itemize}

\paragraph*{B.1.2 Formal Assumptions}\label{b.1.2-formal-assumptions}
\addcontentsline{toc}{paragraph}{B.1.2 Formal Assumptions}

\subparagraph*{Assumption 1: System Regularity (Differentiability and
Strict
Convexity)}\label{assumption-1-system-regularity-differentiability-and-strict-convexity}
\addcontentsline{toc}{subparagraph}{Assumption 1: System Regularity
(Differentiability and Strict Convexity)}

The global loss function \(\mathcal{L}_{\text{global}}(\mathbf{X})\) is
defined as the sum of system loss and all individual costs:
\(\mathcal{L}_{\text{global}}(\mathbf{X}) = \mathcal{L}_{\text{system}}(\mathbf{X}) + \sum_{j=1}^{N} C_j(\mathbf{x}_j)\).

\begin{itemize}
\tightlist
\item
  \(C^2\) Continuity: \(\mathcal{L}_{\text{global}}(\mathbf{X})\) is
  twice continuously differentiable (\(C^2\)) with respect to all agent
  actions \(\mathbf{X} \in \mathcal{X}\). This ensures the existence of
  the incentive gradient \(\mathbf{G}_i\) and the second-order
  conditions necessary for stability.
\item
  Strict Convexity: \(\mathcal{L}_{\text{global}}(\mathbf{X})\) is
  strictly convex jointly in \(\mathbf{X}\) on \(\mathcal{X}\). A global
  minimum exists, by continuity when \(\mathcal{X}\) is compact and by
  assumption when \(\mathcal{X} = \mathbb{R}^{Nk}\); strict convexity
  makes it unique, and we denote it \(\mathbf{X}^*\). Strict convexity
  of each \(C_i\) does not by itself imply this property, since
  \(\mathcal{L}_{\text{system}}\) may couple the actions of different
  agents.
\item
  Note: The computational-cost term \(\kappa(T)\) affects only the
  agent's local optimization (Theorem 2) and does not enter the global
  loss function \(\mathcal{L}_{\text{global}}\).
\end{itemize}

\subparagraph*{Assumption 2: Agent Regularity (Convexity and
Concavity)}\label{assumption-2-agent-regularity-convexity-and-concavity}
\addcontentsline{toc}{subparagraph}{Assumption 2: Agent Regularity
(Convexity and Concavity)}

The individual agent's decision problem must be well-behaved to ensure a
unique optimum exists for the agent's utility maximization step.

\begin{itemize}
\tightlist
\item
  Cost Convexity: The agent's cost function, \(C_i(\mathbf{x}_i)\), is
  convex and \(C^2\) continuous in \(\mathbf{x}_i\). Cost
  \(C_i(\mathbf{x}_i)\) is agent-specific and depends only on
  \(\mathbf{x}_i\), not on \(\mathbf{X}_{-i}\).
\item
  Utility Concavity: Because \(U_i\) and
  \(-\mathcal{L}_{\text{global}}\) share the same gradient with respect
  to \(\mathbf{x}_i\), and since \(\mathcal{L}_{\text{global}}\) is
  strictly convex in \(\mathbf{x}_i\) for every fixed
  \(\mathbf{X}_{-i}\) by Assumption 1, \(U_i\) is strictly concave in
  \(\mathbf{x}_i\). This yields a unique best response for any
  \(\mathbf{X}_{-i}\) in the forward pass.
\end{itemize}

\subparagraph*{Remark B.1 (Incentive Field
Integrability)}\label{remark-b.1-incentive-field-integrability}
\addcontentsline{toc}{subparagraph}{Remark B.1 (Incentive Field
Integrability)}

The incentive field
\(\mathbf{G}_i = -\nabla_{\mathbf{x}_i}\mathcal{L}_{\text{global}}\) is
the gradient of a \(C^2\) function on a convex domain and is therefore
conservative: its line integral along any path of unilateral deviation
by \(A_i\) depends only on the endpoints. This follows from the
definition of \(\mathbf{G}_i\) and Assumption 1 and requires no
additional assumption; in particular, it does not depend on the D-DAG
topology. Since the transfer \(t_i\) is defined directly as a scalar
(B.1.1), no integration of \(\mathbf{G}_i\) is required. The topology
matters instead for how the price is computed, by backpropagation
through the graph (Appendix C.12).

\subsubsection*{B.2 Fundamental Lemmas for the
DPM}\label{b.2-fundamental-lemmas-for-the-dpm}
\addcontentsline{toc}{subsubsection}{B.2 Fundamental Lemmas for the DPM}

The following properties are required for the DPM to exist and function.
Let
\(\mathbf{X} = (\mathbf{x}_1, \ldots, \mathbf{x}_N) \in \mathcal{X}\) be
the vector of actions, and \(\mathbf{X}_{-i}\) be the actions of all
agents excluding \(A_i\).

\paragraph*{Lemma 1: Marginal Payoff
Equivalence}\label{lemma-1-marginal-payoff-equivalence}
\addcontentsline{toc}{paragraph}{Lemma 1: Marginal Payoff Equivalence}

At the optimum, the DPM price equals each agent's marginal cost.

\emph{Lemma 1:} At the optimal action \(\mathbf{x}_i^*\), defined as the
\(i\)-th component of \(\mathbf{X}^*\), the marginal transfer received
by the agent is equal to the marginal cost of its action.

\emph{Proof:} An agent \(A_i\) maximizes its utility
\(U_i(\mathbf{x}_i, \mathbf{X}_{-i}) = t_i(\mathbf{x}_i, \mathbf{X}_{-i}) - C_i(\mathbf{x}_i)\)
(B.1.1). Since \(\mathbf{X}^*\) is an interior minimizer of
\(\mathcal{L}_{\text{global}}\) (Assumption 1),
\(\nabla_{\mathbf{x}_i}\mathcal{L}_{\text{global}}(\mathbf{X}^*) = 0\),
so the first-order condition (FOC) holds at \(\mathbf{X}^*\):

\[\nabla_{\mathbf{x}_i} U_i \Big|_{\mathbf{X}^*} = \mathbf{G}_i(\mathbf{X}^*) = 0 \tag{B.1}\]
\[\nabla_{\mathbf{x}_i} t_i - \nabla_{\mathbf{x}_i} C_i(\mathbf{x}_i) = 0 \implies -\nabla_{\mathbf{x}_i} \mathcal{L}_{\text{system}}(\mathbf{X}^*) = \nabla_{\mathbf{x}_i} C_i(\mathbf{x}_i^*) \tag{B.2}\]

This condition states that the marginal transfer, the price the DPM
attaches to the agent's effect on the system loss, must equal the
marginal cost of the action at the optimum. Both sides of (B.2) are
vectors in \(\mathbb{R}^k\). For
\(\mathbf{X}_{-i} \neq \mathbf{X}^*_{-i}\), the agent's best response
may lie on the boundary of \(\mathcal{X}_i\), where (B.2) need not hold.

\subsubsection*{B.3 Theorem 1: Groves-Form Transfers and Potential-Game
Alignment}\label{b.3-theorem-1-groves-form-transfers-and-potential-game-alignment}
\addcontentsline{toc}{subsubsection}{B.3 Theorem 1: Groves-Form
Transfers and Potential-Game Alignment}

This is the central result: given the other agents' actions, maximizing
utility is equivalent to minimizing the global loss in \(\mathbf{x}_i\).

\emph{Theorem 1 (Groves-Form Transfers and Potential-Game Alignment):}
For any agent \(A_i\) with action \(\mathbf{x}_i \in \mathcal{X}_i\) and
under the regularity conditions in B.1, the agent's utility maximization
under the DPM transfer \(t_i\) is mathematically equivalent to
minimizing the global loss with respect to \(\mathbf{x}_i\), for every
\(\mathbf{X}_{-i}\). The DPM therefore induces an exact potential game
with potential \(-\mathcal{L}_{\text{global}}\), whose unique Nash
equilibrium is the global optimum \(\mathbf{X}^*\).

\paragraph*{System Definitions}\label{system-definitions}
\addcontentsline{toc}{paragraph}{System Definitions}

The system's objective is to find the global optimum by minimizing the
global loss (\(\mathcal{L}_{\text{global}}\)):

\[\mathcal{L}_{\text{global}}(\mathbf{X}) = \mathcal{L}_{\text{system}}(\mathbf{X}) + \sum_{j=1}^{N} C_j(\mathbf{x}_j) \tag{B.3}\]

The individual agent's objective is to maximize its local utility
(B.1.1):

\[U_i(\mathbf{x}_i, \mathbf{X}_{-i}) = t_i(\mathbf{x}_i, \mathbf{X}_{-i}) - C_i(\mathbf{x}_i), \qquad t_i(\mathbf{x}_i, \mathbf{X}_{-i}) = -\mathcal{L}_{\text{system}}(\mathbf{x}_i, \mathbf{X}_{-i}) + h_i(\mathbf{X}_{-i}) \tag{B.4}\]

\paragraph*{DPM Incentive and Equivalence
Proof}\label{dpm-incentive-and-equivalence-proof}
\addcontentsline{toc}{paragraph}{DPM Incentive and Equivalence Proof}

The DPM transmits the marginal system price
\(-\nabla_{\mathbf{x}_i}\mathcal{L}_{\text{system}}\). Subtracting its
own marginal cost, agent \(A_i\) obtains the incentive signal, which
represents its net marginal utility. Because no cost function \(C_j\)
with \(j \neq i\) depends on \(\mathbf{x}_i\):

\[\mathbf{G}_i = \nabla_{\mathbf{x}_i} U_i = - \nabla_{\mathbf{x}_i} \mathcal{L}_{\text{global}} \tag{B.5}\]

Substituting (B.3) into (B.4) expresses the agent's utility in terms of
the global loss:

\[U_i(\mathbf{x}_i, \mathbf{X}_{-i}) = -\mathcal{L}_{\text{global}}(\mathbf{X}) + h_i(\mathbf{X}_{-i}) + \sum_{j \neq i} C_j(\mathbf{x}_j) \tag{B.6}\]

The private cost \(C_i\) is already contained inside
\(\mathcal{L}_{\text{global}}\) and is not subtracted a second time.
Since \(t_i\) is defined as a scalar, no integration of \(\mathbf{G}_i\)
is required (Remark B.1): taking the gradient of (B.6) in
\(\mathbf{x}_i\) recovers (B.5). For any unilateral deviation from
\(\mathbf{x}_i\) to \(\mathbf{x}_i'\), the last two terms of (B.6) are
unchanged, so

\[U_i(\mathbf{x}_i', \mathbf{X}_{-i}) - U_i(\mathbf{x}_i, \mathbf{X}_{-i}) = -\big[\mathcal{L}_{\text{global}}(\mathbf{x}_i', \mathbf{X}_{-i}) - \mathcal{L}_{\text{global}}(\mathbf{x}_i, \mathbf{X}_{-i})\big] \tag{B.7}\]

Hence

\[\underset{\mathbf{x}_i}{\operatorname{argmax}}\, U_i(\mathbf{x}_i, \mathbf{X}_{-i}) = \underset{\mathbf{x}_i}{\operatorname{argmax}} \Big( -\mathcal{L}_{\text{global}}(\mathbf{X}) + h_i(\mathbf{X}_{-i}) + \sum_{j \neq i} C_j(\mathbf{x}_j) \Big) \tag{B.8}\]

and, because \(h_i(\mathbf{X}_{-i})\) and
\(\sum_{j \neq i} C_j(\mathbf{x}_j)\) do not depend on \(\mathbf{x}_i\),
the optimization problem is equivalent to minimizing the global loss
given \(\mathbf{X}_{-i}\):

\[\underset{\mathbf{x}_i}{\operatorname{argmax}}\, U_i(\mathbf{x}_i, \mathbf{X}_{-i}) = \underset{\mathbf{x}_i}{\operatorname{argmin}}\, \mathcal{L}_{\text{global}}(\mathbf{x}_i, \mathbf{X}_{-i}) \quad \text{for every } \mathbf{X}_{-i} \tag{B.9}\]

By (B.7), the induced game is an exact potential game with potential
\(-\mathcal{L}_{\text{global}}\) \citep{monderer1996potential}. Since
\(\mathcal{L}_{\text{global}}\) is convex and differentiable and
\(\mathcal{X}\) is a product of convex sets, any profile at which no
agent can lower \(\mathcal{L}_{\text{global}}\) via a unilateral
deviation is a global minimizer. By strict convexity (Assumption 1),
this minimizer is unique, and the unique Nash equilibrium is therefore
\(\mathbf{X}^*\).

This structural alignment shows that the DPM transfer mirrors the Groves
form: under separable costs, \(t_i\) differs from
\(-[\mathcal{L}_{\text{system}} + \sum_{j \neq i} C_j] + h_i(\mathbf{X}_{-i})\)
only by a term in \(\mathbf{X}_{-i}\), where VCG constitutes the special
case utilizing the Clarke pivot \citep{clarke1971multipart}. The DPM is
thus a differentiable counterpart of the VCG principle in action spaces.
Because agents select continuous actions directly rather than reporting
hidden types, and because each best response depends on
\(\mathbf{X}_{-i}\), this formulation guarantees equilibrium alignment
rather than dominant-strategy incentive compatibility. Like any Groves
mechanism, the DPM also gives up budget balance: total transfers
\(\sum_i t_i = -N\mathcal{L}_{\text{system}} + \sum_i h_i(\mathbf{X}_{-i})\)
need not sum to zero, so the Planner must fund or absorb the difference
\citep{green1977characterization}. With \(h_i \equiv 0\) and a
non-negative system loss, agents are charged for the system loss rather
than paid; the terms \(h_i\) determine the net direction and size of
payments. Transfers are settled once, on the realized outcome; during
the iterations the Planner communicates only their derivative, the
price, which is all an agent needs to maximize \(t_i - C_i\). The terms
\(h_i\) do not affect incentives; given an upper bound on agents' costs,
they can be chosen to secure voluntary participation, at the cost of a
larger budget. A type-reporting variant yielding dominant-strategy
guarantees, and the design of \(h_i\) for budget and participation, are
reserved for later investigation.

\subsubsection*{B.4 Theorem 2: Formal Satisficing
Condition}\label{b.4-theorem-2-formal-satisficing-condition}
\addcontentsline{toc}{subsubsection}{B.4 Theorem 2: Formal Satisficing
Condition}

Theorem 2 specifies the bounded rationality of the MBI agent by defining
the optimal stopping condition for continuous computational search
effort.

\emph{Theorem 2 (Formal Satisficing Condition):} The MBI agent,
operating under Simon's BR and searching by gradient ascent on its
utility, halts its search at the effort level \(T^*\) where the marginal
incentive gain is outweighed by the marginal cost of effort. If
\(\kappa(T)\) is convex, \(T^*\) maximizes the agent's utility net of
effort cost, and the incentive remaining at \(T^*\) is bounded by the
marginal effort cost (B.14).

This captures Simon's satisficing principle in continuous time: the
agent halts updating once the net marginal utility gain falls below an
aspiration threshold set by its internal effort cost. Formally, the rule
is optimal stopping under search costs \citep{stigler1961economics}; the
aspiration level is the residual bound (B.14).

\paragraph*{Formalizing the Bounded Rationality
Constraint}\label{formalizing-the-bounded-rationality-constraint}
\addcontentsline{toc}{paragraph}{Formalizing the Bounded Rationality
Constraint}

The MBI agent operates under BR, formalized by treating the agent's
action \(\mathbf{x}_i\) as an explicit function of the computational
effort \(T\), with the joint profile \(\mathbf{X}_{-i}\) held fixed
during the search trajectory:

\[\mathbf{x}_i(T): [0, \infty) \to \mathcal{X}_i \tag{B.10}\]

The parameter \(T\) represents continuous resource metrics such as
search depth, internal convergence tolerance, or local gradient steps.

Two costs must be kept apart: the economic cost of an action,
\(C_i(\mathbf{x}_i)\) (which alters \(\mathcal{L}_{\text{global}}\) and
Theorem 1), and the computational effort cost, \(\kappa(T)\), which
measures the overhead of the local optimization search itself.

Conditions on the Search Trajectory \(\mathbf{x}_i(T)\):

\begin{enumerate}
\def\labelenumi{\arabic{enumi}.}
\tightlist
\item
  Differentiability: The agent's action path \(\mathbf{x}_i(T)\) is
  continuously differentiable with respect to \(T\) and remains in the
  interior of \(\mathcal{X}_i\). This ensures the existence and
  continuity of the internal derivative
  \(\frac{\partial \mathbf{x}_i}{\partial T}\), validating the
  multi-variable chain rule.
\item
  Gradient Ascent: The agent searches along its incentive direction,
  \(\frac{\partial \mathbf{x}_i}{\partial T} = \eta \, \mathbf{G}_i(\mathbf{x}_i(T))\)
  with step size \(\eta > 0\). Each unit of search effort weakly
  increases \(U_i\), since
  \(\mathbf{G}_i \cdot \frac{\partial \mathbf{x}_i}{\partial T} = \eta \|\mathbf{G}_i\|^2 \geq 0\).
  Here \(\eta\) is the agent's search rate, the progress per unit of
  effort; in Theorem 4, one unit of effort is one iteration, so \(\eta\)
  also becomes the discrete step size, and only there is it constrained
  by the stability condition \(\eta < 2/L_{\nabla}\).
\end{enumerate}

The agent maximizes its net local objective over effort \(T\):

\[U(T) = U_i(\mathbf{x}_i(T), \mathbf{X}_{-i}) - \kappa(T) \tag{B.11}\]

The agent halts its search and satisfices at effort level \(T^*\) when
the marginal utility gain with respect to effort becomes non-positive
(\(\frac{\partial U}{\partial T} \big| _ {T=T^*} \leq 0\)). Applying the
chain rule and the identity \(\mathbf{G}_i = \nabla_{\mathbf{x}_i} U_i\)
(B.5) yields:

\[\frac{\partial U}{\partial T} = \mathbf{G}_i \cdot \frac{\partial \mathbf{x}_i}{\partial T} - \frac{\partial \kappa}{\partial T} \tag{B.12}\]

The formal satisficing condition is defined as:

\[\mathbf{G}_i \cdot \frac{\partial \mathbf{x}_i}{\partial T} \bigg|_{T=T^*} \leq \frac{\partial \kappa(T)}{\partial T} \bigg|_{T=T^*} \tag{B.13}\]

Under the gradient-ascent condition, the left side of (B.13) evaluates
to \(\eta \|\mathbf{G}_i\|^2\). Its derivative along the search path is
\(2\eta^2 \, \mathbf{G}_i^{\top} \nabla^2_{\mathbf{x}_i} U_i \, \mathbf{G}_i \leq 0\),
because \(U_i\) is strictly concave in \(\mathbf{x}_i\) by Assumption 2.
The marginal search gain is therefore non-increasing in \(T\). When
\(\kappa(T)\) is convex, its marginal cost is non-decreasing, rendering
\(U(T)\) concave; thus, the first instance of \(T^*\) satisfying (B.13)
globally maximizes \(U(T)\). At \(T^*\), evaluating (B.13) establishes
the residual bound:

\[\|\mathbf{G}_i(\mathbf{x}_i(T^*))\| \leq \sqrt{\frac{1}{\eta} \frac{\partial \kappa}{\partial T} \bigg|_{T=T^*}} \tag{B.14}\]

Whenever computational effort is costly
(\(\frac{\partial \kappa}{\partial T} > 0\)), the agent stops before
reaching its unconstrained best response, leaving a residual incentive
bound determined by the ratio of marginal effort cost to the local
optimization step size.

\subsubsection*{B.5 Theorem 3: Private Costs under Asymmetric
Information}\label{b.5-theorem-3-private-costs-under-asymmetric-information}
\addcontentsline{toc}{subsubsection}{B.5 Theorem 3: Private Costs under
Asymmetric Information}

Theorem 3 examines hidden private costs under asymmetric information,
showing that the equilibrium does not depend on the Planner's knowledge
of costs, and identifying the best estimate when the Planner prices a
cost itself.

\emph{Theorem 3 (Private Costs and Certainty Equivalence):} Let
Assumptions 1 and 2 hold for the true costs. For Part 2, let each cost
\(C_i\) be affine in a private parameter \(\lambda_i\), drawn from a
prior known to the Planner, and let Assumption 1 hold for every
\(\lambda_i\) in the support of the prior.

\begin{enumerate}
\def\labelenumi{\arabic{enumi}.}
\tightlist
\item
  Under the transfer rule defined in B.1.1, the unique Nash equilibrium
  is the social optimum \(\mathbf{X}^*\) for the true costs, for every
  realization of \(\lambda_i\) and every belief the Planner holds about
  it; asymmetric information causes no welfare loss.
\item
  If instead the Planner prices agent \(A_i\)'s cost itself, using a
  point estimate \(\hat{\lambda}_i\) in place of \(\lambda_i\) (as
  evaluated in Appendix C.6), the estimate that minimizes the expected
  global loss is \(\hat{\lambda}_i = E[\lambda_i]\).
\end{enumerate}

\paragraph*{Proof}\label{proof}
\addcontentsline{toc}{paragraph}{Proof}

\emph{Part 1.} The transfer
\(t_i = -\mathcal{L}_{\text{system}} + h_i(\mathbf{X}_{-i})\) contains
no private parameter, and agent \(A_i\) evaluates \(U_i = t_i - C_i\)
using its true private cost realization. Theorem 1 applies to the true
global loss for every realization of \(\lambda_i\), and the unique Nash
equilibrium remains the corresponding social optimum \(\mathbf{X}^*\),
eliminating the requirement to elicit private information. This concerns
the equilibrium and sequential best-response dynamics; simultaneous
gradient dynamics additionally require a known upper bound on the
curvature of the private costs (for affine costs as in Theorem 3, an
upper bound on the support of \(\lambda_i\)) to set the step size
(Section B.6), and may diverge if a true curvature exceeds it (Appendix
C.5).

\emph{Part 2.} Write
\(\mathcal{L}_{\text{global}}(\mathbf{X}; \lambda_i)\) to make the
structural dependence on \(\lambda_i\) explicit, and let
\(\mathbf{X}(\hat{\lambda}_i)\) denote the allocation realized when the
Planner prices with the estimate \(\hat{\lambda}_i\), which is the
minimizer of
\(\mathcal{L}_{\text{global}}(\mathbf{X}; \hat{\lambda}_i)\). Because
\(\mathcal{L}_{\text{global}}\) is affine in \(\lambda_i\), the
expectation over the prior on \(\lambda_i\) for any fixed joint action
profile \(\mathbf{X}\) satisfies:

\[E_{\lambda_i}\big[\mathcal{L}_{\text{global}}(\mathbf{X}; \lambda_i)\big] = \mathcal{L}_{\text{global}}(\mathbf{X}; E[\lambda_i]) \tag{B.15}\]

Hence, for any point estimate \(\hat{\lambda}_i\), the expected global
loss evaluates to:
\[E_{\lambda_i}\big[\mathcal{L}_{\text{global}}(\mathbf{X}(\hat{\lambda}_i); \lambda_i)\big] = \mathcal{L}_{\text{global}}(\mathbf{X}(\hat{\lambda}_i); E[\lambda_i])\]
This term is minimized when the mapping \(\mathbf{X}(\hat{\lambda}_i)\)
minimizes the objective evaluated at the expected value,
\(\mathcal{L}_{\text{global}}(\cdot\,; E[\lambda_i])\), which holds when
\(\hat{\lambda}_i = E[\lambda_i]\). Because the right side of (B.15)
constitutes an average of strictly convex functions, this global
minimizer is unique.

The guarantee established in Part 2 is ex ante: for any specific
realization of \(\lambda_i\), the realized loss can exceed the true
social optimum, as empirically demonstrated in Appendix C.6. Part 1
establishes that direct knowledge of \(\lambda_i\) is unnecessary for
the DPM; Part 2 applies exclusively when the Planner, rather than the
decentralized agent, internalizes \(C_i\). Because neither formulation
relies on explicit message reports, no type-level incentive
compatibility is asserted here. Extending MBI to type-reporting variants
utilizing expected-externality transfers
\citep{daspremont1979incentives} or dominant-strategy mechanisms is
deferred to subsequent study.

\subsubsection*{B.6 Theorem 4: Convergence Under IC and
BR}\label{b.6-theorem-4-convergence-under-ic-and-br}
\addcontentsline{toc}{subsubsection}{B.6 Theorem 4: Convergence Under IC
and BR}

This theorem gives convergence of the joint dynamics, and the residual
error when search is costly.

\emph{Theorem 4 (Convergence Under IC and BR):} Let \(\mathbf{X}^{(t)}\)
be the joint action profile at iteration \(t\), and let all agents
update simultaneously: agent \(A_i\) takes the step
\(\mathbf{x}_i^{(t+1)} = \mathbf{x}_i^{(t)} + \eta\, \mathbf{G}_i(\mathbf{X}^{(t)})\)
if its marginal gain from one step exceeds its marginal effort cost,
\(\eta \|\mathbf{G}_i(\mathbf{X}^{(t)})\|^2 > \frac{\partial \kappa}{\partial T}\)
(B.13), and otherwise satisfices,
\(\mathbf{x}_i^{(t+1)} = \mathbf{x}_i^{(t)}\). Assume the conditions of
Theorem 1, a constant marginal effort cost
\(\frac{\partial \kappa}{\partial T} \geq 0\), and an admissible step
size \(0 < \eta < 2/L_{\nabla}\).

Then:

\begin{enumerate}
\def\labelenumi{\arabic{enumi}.}
\tightlist
\item
  If \(\frac{\partial \kappa}{\partial T} = 0\), the sequence
  \(\{\mathbf{X}^{(t)}\}\) converges to the unique global minimizer
  \(\mathbf{X}^*\).
\item
  If \(\frac{\partial \kappa}{\partial T} > 0\),
  \(\mathcal{L}_{\text{global}}\) is non-increasing along the sequence,
  and the process comes to rest after finitely many updates at a profile
  where every agent satisfies the residual bound (B.14),
  \(\|\mathbf{G}_i\| \leq \sqrt{\frac{1}{\eta}\frac{\partial \kappa}{\partial T}}\).
  These rest profiles approach \(\mathbf{X}^*\) as
  \(\frac{\partial \kappa}{\partial T} \to 0\).
\end{enumerate}

\paragraph*{Sufficient Conditions for
Convergence}\label{sufficient-conditions-for-convergence}
\addcontentsline{toc}{paragraph}{Sufficient Conditions for Convergence}

\begin{enumerate}
\def\labelenumi{\arabic{enumi}.}
\tightlist
\item
  Strict Convexity: \(\mathcal{L}_{\text{global}}\) is strictly convex
  jointly in \(\mathbf{X}\) (Assumption 1). Theorem 4 concerns the
  unconstrained case \(\mathcal{X}_i = \mathbb{R}^k\), as for the losses
  in Appendix C: \(\mathcal{L}_{\text{global}}\) has these properties on
  all of \(\mathbb{R}^{Nk}\) and a minimizer, so its sublevel sets are
  compact and the iterates require no projection. With compact
  \(\mathcal{X}_i\), the gradient steps must be replaced by projected
  gradient steps, whose analysis remains an open direction. Strict
  convexity of individual agent cost functions alone does not guarantee
  joint strict convexity, due to coupling across
  \(\mathcal{L}_{\text{system}}\).
\item
  Lipschitz Continuity of the Gradient and Step Size:
  \(\nabla \mathcal{L}_{\text{global}}\) is Lipschitz continuous with
  constant \(L_{\nabla}\), and the step size satisfies
  \(0 < \eta < 2/L_{\nabla}\) to prevent overshooting. For objectives
  evaluated as sums over agents, such as
  \((\sum_i \mathbf{x}_i - Y^*)^2 + \sum_i \lambda_i \mathbf{x}_i^2\),
  \(L_{\nabla}\) scales linearly with \(N\), which causes the maximum
  admissible step size to contract as \(1/N\).
\end{enumerate}

\paragraph*{Proof}\label{proof-1}
\addcontentsline{toc}{paragraph}{Proof}

In Theorem 4, the satisficing check (B.13) is re-evaluated at every
iteration: an agent that does not update at iteration \(t\) may update
again once the other agents' actions change its incentive. Theorem 2
describes a single search episode with \(\mathbf{X}_{-i}\) fixed; here,
rest means that no agent updates. One iteration corresponds to a unit of
search effort, \(\Delta T = 1\), so the step
\(\mathbf{x}_i^{(t+1)} - \mathbf{x}_i^{(t)} = \eta\,\mathbf{G}_i\) is
the unit-step discretization of the search path of Theorem 2.

Because
\(\nabla_{\mathbf{x}_i}\mathcal{L}_{\text{global}} = -\mathbf{G}_i\)
(B.5), the standard descent inequality for functions with an
\(L_{\nabla}\)-Lipschitz continuous gradient
\citep{nesterov2004introductory} satisfies, where the summation runs
over the subset of agents executing updates at iteration \(t\):

\[\mathcal{L}_{\text{global}}(\mathbf{X}^{(t+1)}) \leq \mathcal{L}_{\text{global}}(\mathbf{X}^{(t)}) - \eta\Big(1 - \frac{\eta L_{\nabla}}{2}\Big) \sum_{i \text{ updating}} \|\mathbf{G}_i(\mathbf{X}^{(t)})\|^2 \tag{B.16}\]

For \(0 < \eta < 2/L_{\nabla}\) the factor in parentheses is strictly
positive, rendering \(\mathcal{L}_{\text{global}}\) non-increasing. The
generated iterates remain confined within the sublevel set
\(\{\mathbf{X} : \mathcal{L}_{\text{global}}(\mathbf{X}) \leq \mathcal{L}_{\text{global}}(\mathbf{X}^{(0)})\}\),
which is compact because a convex function on \(\mathbb{R}^{Nk}\) whose
minimizer set is nonempty and compact (here the single point
\(\mathbf{X}^*\)) has bounded sublevel sets.

\emph{Part 1.} When \(\frac{\partial \kappa}{\partial T} = 0\), every
agent with a non-zero gradient vector \(\mathbf{G}_i \neq 0\) updates.
Summing the descent inequality (B.16) over all past iterations \(t\),
and noting that \(\mathcal{L}_{\text{global}}\) is bounded below by
\(\mathcal{L}_{\text{global}}(\mathbf{X}^*)\), yields
\(\sum_t \|\nabla \mathcal{L}_{\text{global}}(\mathbf{X}^{(t)})\|^2 < \infty\).
So the gradient tends to zero. Every limit point constitutes a
stationary point, which by joint strict convexity maps uniquely to the
social optimum \(\mathbf{X}^*\). Since the action profiles populate a
compact domain, \(\mathbf{X}^{(t)} \to \mathbf{X}^*\), satisfying:

\[\nabla_\mathbf{X} \mathcal{L}_{\text{global}}(\mathbf{X}^*) = 0 \tag{B.17}\]

\emph{Part 2.} When \(\frac{\partial \kappa}{\partial T} > 0\), an agent
updates its action if and only if
\(\|\mathbf{G}_i\|^2 > \frac{1}{\eta}\frac{\partial \kappa}{\partial T}\).
By (B.16), every iteration containing at least one active agent update
reduces the objective function by an amount strictly greater than
\((1 - \frac{\eta L_{\nabla}}{2})\frac{\partial \kappa}{\partial T}\).
Because \(\mathcal{L}_{\text{global}}\) is bounded below, the system can
only execute a finite number of such updates before coming to rest. Once
updates cease, every agent satisfies the residual bound (B.14). The
compact set of resting profiles contains \(\mathbf{X}^*\) and contracts
toward the true social optimum as
\(\frac{\partial \kappa}{\partial T} \to 0\) due to the continuity of
\(\nabla \mathcal{L}_{\text{global}}\) and the uniqueness of its root.

At \(\mathbf{X}^*\), the marginal transfer received by each agent equals
the marginal cost of its action (Lemma 1), and by Theorem 1 no agent has
an incentive to deviate. The update rule matters: simultaneous exact
best responses need not converge even in potential games, whereas
simultaneous gradient steps with \(\eta < 2/L_{\nabla}\) form a
gradient-descent iteration on \(\mathcal{L}_{\text{global}}\) and
therefore converge \citep{bertsekas1999nonlinear}.

Gradient steps model the agents' bounded search (Theorem 2) rather than
an imposed algorithm: an agent able to compute its exact best response
may do so, and the equilibrium of Theorem 1 is unchanged. If agents
best-respond one at a time, the iteration is block coordinate descent on
the strictly convex \(\mathcal{L}_{\text{global}}\) and converges
\citep{bertsekas1999nonlinear}; with simultaneous updates, the common
step size \(\eta < 2/L_{\nabla}\) must be announced by the Planner,
which requires a known bound on the curvature of the private costs.

The satisficing check (B.13) uses the continuous-effort marginal gain
\(\eta \|\mathbf{G}_i\|^2\); the gain from one discrete step by the
agent alone, holding the other agents' actions fixed, is at least
\(\eta(1 - \eta L_{\nabla}/2)\|\mathbf{G}_i\|^2\), so an agent may take
a final step worth slightly less than its effort cost.

\subsection*{C: Empirical Validation}\label{c-empirical-validation}
\addcontentsline{toc}{subsection}{C: Empirical Validation}

Table 3 is structured as follows. The full experimental sections C.1 to
C.12, including methodology, results, and code, appear in the
accompanying notebook in the project repository (Appendix D), with
references to Appendix C.\(n\) pointing to the corresponding notebook
sections. Section C.4, a comparison with reward-based learning, is
included in the notebook as an illustration and is not used in the
paper. Sections C.1 to C.11 test system losses where agents interact
through the sum of their outputs, while C.12 tests pricing across a
compositional D-DAG. The free-rider benchmark is the Nash equilibrium
without DPM prices, where each agent bears its own cost plus an equal
\(1/N\) share of the system loss; the no-transfer benchmark ignores the
system loss entirely.

\begin{table}[H]
    \centering
    \caption{Summary of Empirical Validation of the MBI Mechanism}
    \begin{tabular}{|m{4.1cm}|c|m{7.4cm}|}
        \hline
        \textbf{Category} & \textbf{Sections} & \textbf{Summary of Results} \\
        \hline \hline
        \textbf{I. Mechanism Fidelity} & C.1, C.2, C.3, C.12 & With costless effort, agents converge to the global optimum to numerical precision, from two agents (C.1) to $N = 100$ symmetrical agents (C.2) and five heterogeneous producers in a stylized economic dispatch (C.3), where the DPM price equalizes all marginal costs. MBI's loss is $9.1$ (C.2), $1.6$ (C.3), and $1.4$ (C.12) times lower than the free-rider Nash equilibrium. In a compositional D-DAG, backpropagation reproduces the chain-rule price exactly (C.12). \\
        \hline
        \textbf{II. Scale} & C.11 & The cost per iteration grows close to linearly in $N$ up to $10^7$ agents, while for summed losses the iterations to converge grow roughly in proportion to $N$ (C.11). \\
        \hline
        \textbf{III. Convergence, Stability and BR} & C.1, C.5, C.8, C.9 & Gradient steps converge for every $\eta < 2/L_{\nabla}$ and diverge above it (C.5). Satisficing agents stop within the residual bound (B.14) (C.1), and the stopping rule (B.13) minimizes total cost up to one discrete step (C.9). In a non-convex landscape, MBI reaches a local minimum $45\%$ better than free-riding but $62\%$ above the global minimum (C.8). \\
        \hline
        \textbf{IV. Information and Robustness} & C.6, C.7, C.10 & At equilibrium, the Planner requires no knowledge of private costs (C.6, Theorem 3 Part 1). If the Planner prices costs itself, a $10\%$ error in all costs raises the loss by $0.001\%$, a threefold underestimate of a single cost by $33\%$, and the expected cost minimizes the expected loss (C.6). Under persistent noise the dynamics remain stable, with a stationary excess loss matching its closed form (C.7). A private, time-varying cost unknown to the Planner is internalized (C.10). \\
        \hline
    \end{tabular}
\end{table}

\subsection*{D: Reproducibility}\label{d-reproducibility}
\addcontentsline{toc}{subsection}{D: Reproducibility}

The source code supporting all simulations and empirical analyses in
this paper is publicly available on the project's GitHub repository at

\texttt{github.com/stevefatz95/mechanism\_based\_intelligence}.

All experiments were conducted using Python 3.10.13, PyTorch 2.9.0, and
NumPy 2.0.1 on a Mac M4 architecture. All random initializations are
explicitly seeded, ensuring that every reported metric is deterministic
and reproducible, excluding the hardware-dependent wall-clock timings
reported in Appendix C.11. The arguments of Theorems 1 to 4 have been
formally checked in Lean 4 with Mathlib, taking as hypotheses the
first-order characterization of differentiable convex functions, the
standard descent inequality for functions with Lipschitz continuous
gradients, a terminal error bound (which holds, for example, under
strong convexity) in place of compactness for Theorem 4, and the
monotonicity of the marginal effort gain derived in Theorem 2. The Lean
source files are available in the same repository.

\end{document}